\documentclass[11pt, a4paper]{article}
\usepackage[round]{natbib}
\usepackage[breaklinks,citecolor=black,urlcolor=black]{hyperref}
\usepackage{booktabs} 
\usepackage[font=small,labelfont=bf]{caption} 
\usepackage{amsfonts, amsmath, amsthm, amssymb}
\usepackage{indentfirst} 
\usepackage{algorithmic}
\usepackage{ascmac}
\usepackage{url}
\usepackage{threeparttable}
\usepackage{breakurl}

\usepackage{graphicx}
\begin{document}
\title{A Bayesian Graphical Approach for Large-Scale Portfolio Management with Fewer Historical Data  \thanks{This research is supported by Keio University Doctoral Student Grant-in-Aid Program.}}\author{Sakae Oya\footnote{\url{sakae.prosperity21@keio.jp}
}\\Graduate School of Economics, Keio University, Tokyo, Japan}
\date{}
\maketitle
\begin{abstract}
Managing a large-scale portfolio with many assets is one of the most challenging tasks in the field of finance.
  It is partly because estimation of either covariance or precision matrix of asset returns tends to be unstable or even infeasible when the number of assets $p$ exceeds the number of observations $n$.
  For this reason, most of the previous studies on portfolio management have focused on the case of $p < n$.
  To deal with the case of $p > n$, we propose to use a new Bayesian framework based on adaptive graphical LASSO for estimating the precision matrix of asset returns in a large-scale portfolio.
  Unlike the previous studies on graphical LASSO in the literature, our approach utilizes a Bayesian estimation method for the precision matrix proposed by \cite{ONGLASSO} so that the positive definiteness of the precision matrix should be always guaranteed.
  As an empirical application, we construct the global minimum variance portfolio of $p=100$ for various values of $n$ with the proposed approach as well as the non-Bayesian graphical LASSO approach, and compare their out-of-sample performance with the equal weight portfolio as the benchmark.
  In this comparison, the proposed approach produces more stable results than the non-Bayesian approach in terms of Sharpe ratio, portfolio composition and turnover.
  Furthermore, the proposed approach succeeds in estimating the precision matrix even if $n$ is much smaller than $p$ and the non-Bayesian approach fails to do so.

\noindent
\\\textbf{Keywords:} Bayesian adaptive graphical LASSO, Global minimum variance portfolio, $p > n$ problem, Positive definiteness, Precision matrix
\end{abstract}

\section{Introduction}
\label{intro}
Since \cite{Markowitz} proposed the mean-variance model for optimal portfolio selection, it has remained an important foundation in research of portfolio theory as well as in practice of portfolio management.
For example, many of robot advisor services developed in the recent fintech boom are basically based on the mean-variance model or its variants such as the Black-Litterman approach (\citeauthor{BlackLitterman1} (\citeyear{BlackLitterman1}, \citeyear{BlackLitterman2})).
Although numerous solutions have been proposed for various problems in the mean-variance model over its long history, one of the most challenging tasks is how to construct the optimal portfolio when the number of assets $p$ exceeds the number of observations of asset returns $n$. This issue is often referred to as the $p > n$ problem in the literature.

Let us restate this problem in the context of optimal portfolio selection.
For the sake of simplicity, we consider the variance minimization problem without setting the target level of the expected return of the portfolio, i.e.,
\begin{equation}
\label{gmvopt}
\begin{split}
 \min_{\boldsymbol{w}} &\quad \boldsymbol{w^{\intercal}} \boldsymbol{\Sigma} \boldsymbol{w} \\
\text{s.t.} &\quad \boldsymbol{\iota^{\intercal}} \boldsymbol{w} = 1,\\
\end{split}
\end{equation}
where
\begin{equation*}
\boldsymbol{w} = \begin{bmatrix} w_{1} \\ \vdots \\ w_{p} \end{bmatrix},
\ \boldsymbol{\Sigma} = \begin{bmatrix} \sigma_{1}^{2} &  \ldots & \sigma_{1p} \\ \vdots &  \ddots & \vdots \\ \sigma_{p1} & \ldots & \sigma_{p}^{2} \end{bmatrix},
\ \boldsymbol{\iota} = \begin{bmatrix} 1 \\ \vdots \\ 1 \end{bmatrix}.
\end{equation*}
$\boldsymbol{w}$ is a $p \times 1$ vector of allocation weights and $\boldsymbol{\Sigma}$ is a $p \times p$ covariance matrix of asset returns. $\boldsymbol{\iota}$ is a $p \times 1$ vector whose elements are all equal to one.
In this setup, the solution of \eqref{gmvopt} is given by
\begin{equation}
\label{gmv}
\boldsymbol{w^{\text{GMV}}} = \frac{1}{\boldsymbol{\iota^{\intercal}}\boldsymbol{\Sigma^{-1}}\boldsymbol{\iota}}\boldsymbol{\Sigma^{-1}}\boldsymbol{\iota},
\end{equation}
where $\boldsymbol{w^{{GMV}}}$ is called the global minimum variance portfolio.

Note that $\boldsymbol{w^{GMV}}$ depends on $\boldsymbol{\Sigma^{-1}}$, the inverse matrix of the covariance matrix $\boldsymbol{\Sigma}$ or the precision matrix.
For later use, we define $\boldsymbol{\Omega} = \boldsymbol{\Sigma^{-1}}$.
Since the precision matrix $\boldsymbol{\Omega}$ is unknown in practice, we need to estimate it with asset return data before we apply the formula in \eqref{gmv}.
The simplest estimation method for $\boldsymbol{\Omega}$ is to replace each element of $\boldsymbol{\Sigma}$ with its sample analog, which is called the sample covariance matrix, and then compute the inverse of the sample covariance matrix to obtain an estimate of $\boldsymbol{\Omega}$.
In principle, however, we cannot compute the inverse of the sample covariance matrix when the number of assets $p$ is larger than the number of asset return observations $n$.
Moreover, even if $p$ is smaller then $n$, it is well known that computation of the inverse of the sample covariance matrix tends to be unstable as $p$ approaches $n$.
Therefore, in order to replace $\boldsymbol{\Sigma}$ in \eqref{gmv} with the sample covariance matrix, we need to prepare a sufficiently larger number of asset return observations than the number of assets in the portfolio.

This requirement would be a great hindrance in practice.
For example, when we want to manage 500 assets in a fund, we need more than two years' worth of data in case of daily returns and more than 42 years' worth of data in case of monthly returns.
The latter is rather impractical, of course.

Let us point out another situation related to the $p > n$ problem. Suppose a large-scale IPO is made due to a privatization of a state-owned enterprise or a large-cap firm merges with another one and relists itself.
For those stocks with large market-cap, available historical data after IPO or merger is too few.
Thus it is difficult to evaluate the risk of a portfolio which includes those stocks because of the $p > n$ problem.

Besides the problem of data availability, using long historical asset return data may cause another problem.
Suppose we observe large market turmoil in the past few days.
Although it is preferable to incorporate this recent shock into the estimation of the precision matrix, influence of older observations well before the shock is still so dominating in long historical data that we could underestimate the risk of the new shock and it might result in under-performance.
To avoid this issue, we may cut out most of the older observations. But then again we are faced with the $p > n$ problem due to a shorten sample period.

In the literature, many researchers have tackled the $p > n$ problem in different contexts and proposed many possible solutions.
Among them, one of the most widely explored methods is dimension compression.
In particular, factor models are quite popular as tools for dimension compression in the field of finance.
A typical factor model assumes that the variation in asset returns is mostly explained by a set of common factors and the residuals are independent of each other as well as the common factors.
In this way, the covariance matrix of asset returns $\boldsymbol{\Sigma}$ is decomposed into two matrices: the covariance matrix induced by the common factors and the diagonal covariance matrix of the residuals.
It is well known that $\boldsymbol{\Sigma}$ in a factor model will be non-singular as long as the common factors are linearly independent.
In a typical factor model, the number of the common factors, say $k$, is much smaller than the number of observations $n$.
For most cases, this property itself is sufficient to guarantee the linear independence among the common factors and the existence of the precision matrix $\boldsymbol{\Omega}$.
Furthermore, since $k < p$, the dimension of the parameter space is largely reduced.
So it is regarded as a type of dimension compression method.
There are many variants of factor models including dynamic ones, but arguably the most famous one is the three-factor model proposed by \cite{Fama_French3}.

As the factor model gains popularity in both academia and business, the number of potential candidates for common factors has been exploding.
According to \cite{HarveyF316}, it reaches 316 and counting.
Thus it is necessary to select appropriate factors among a huge set of candidates in practice.
Here again, we can utilize dimension compression for this purpose.
In the context of linear regression with many explanatory variables, the penalized regression method is widely used for simultaneously selecting appropriate variables and estimating the corresponding coefficients as a convenient way to apply dimension compression.
It includes least absolute shrinkage and selection operator or LASSO (\cite{Tibshirani}), elastic net (\cite{Zou_and_Hastie}), adaptive LASSO (\cite{HuiZou2006}), Bayesian LASSO (\cite{Park_and_Casella}), horseshoe prior (\citeauthor{CPS1} (\citeyear{CPS1}, \citeyear{CPS2})), Bayesian adaptive LASSO (\cite{Alhamzawi}, \cite{Lengetal}), generalized double Pareto (\cite{ADL}) among others.
In relation to the $p > n$ problem, \cite{BCR} proposed a new approach called ``Bayesian compressed regression'' which randomly compresses a scaled predictor vector prior to analysis.
\cite{BCR} tested a case of $p = 25,000$ and $n = 110$ in simulation, though it has not yet been applied to asset management as far as we know.

Another popular approach\footnote{In addition to these two major approaches we mention here, alternative approaches such as \cite{LW2004}, \cite{RMT99} among others are known in the literature.} is to use a Gaussian graphical model\footnote{Although we separetely explain the factor model and the Gaussian graphical model in this introduction, distinction between them is rather arbitrary. In practice, we can combine both approaches together as we do in Section 3.} for direct estimation of the precision matrix $\boldsymbol{\Omega}$.
The Gaussian graphical model take advantage of the fact that all elements in $\boldsymbol{\Omega}$ can be treated as unknown parameters in the likelihood function if asset returns are supposed to jointly follow a $p$-dimensional multivariate normal distribution with the zero mean vector and the covariance matrix $\boldsymbol{\Sigma}$.
The term ``graphical'' comes from a property of the multivariate normal distribution that any pair of normal random variables are independent if and only if the corresponding off-diagonal element in $\boldsymbol{\Omega}$ is zero.
Therefore $\boldsymbol{\Omega}$ gives a network of dependence among the random variables where non-zero off-diagonal elements are regarded as links connecting random variables (nodes in the context of graphical modeling) and any zero off-diagonal element indicates no link between two nodes.

Since the number of elements in $\boldsymbol{\Omega}$ is $p(p+1)/2$, the number of parameters to be estimated will be considerably high for a large-scale graphical model.
Therefore it is important to force weak and unessential links to be zero so that the estimated structure of network would become more sparse and interpretable.
To penalize inclusion of such redundant links in the model, aforementioned LASSO proposed by \cite{Tibshirani} has been applied to the Gaussian graphical model by \cite{Meinshausen}, \cite{Friedman}, \cite{Yuan_and_Lin}, \cite{Banerjee}, \cite{Guo} among others. This type of LASSO is called graphical LASSO or glasso. Alternatively, \cite{Finegold} proposed tlasso which used the multivariate Student's t distribution in place of the multivariate normal distribution in the likelihood function.

In this paper, to deal with the $p > n$ problem in large-scale portfolio management, we pursue the second approach and propose to utilize graphical LASSO to cull unnecessary dependence among assets in $\boldsymbol{\Omega}$ so that the global minimum variance portfolio \eqref{gmv} should be stabilized even in case of $p>n$.
The Gaussian graphical model was already applied to portfolio optimization by \cite{Goto}, \cite{Brownelees}, \cite{Torri} among others.
Especially, \cite{Torri} examined performance of the global minimum variance portfolio \eqref{gmv} constructed by both glasso and tlasso in long-term asset management, though \cite{Torri} limited the scope of their study in case of $p < n$.
Instead, we try to construct \eqref{gmv} in case of $p > n$ and push the envelope of graphical LASSO.
As far as we know, ours is the first attempt to estimate $\boldsymbol{\Omega}$ in case of $p >  n$ and use it in performance comparison of long-term asset management.

For this purpose, we develop a data-driven portfolio framework based on a Bayesian version of graphical LASSO.
From the Bayesian perspective, graphical LASSO is regarded as a maximum a posteriori (MAP) estimator with the Laplace prior for each element in $\boldsymbol{\Omega}$. Therefore it is natural for Bayesian statisticians to extend graphical LASSO by introducing a hierarchical structure among priors such as adaptive graphical LASSO (\cite{Wang}) and covariance LASSO (\cite{Khondker}), or replacing the Laplace prior with alternative shrinkage priors such as the spike-and-slab prior (\cite{wang2015}) and the horseshoe prior (\cite{Li_Craig_Bhadra}).
These aforementioned previous studies on Bayesian graphical LASSO except for \cite{Khondker} relies on the block Gibbs sampler proposed by \cite{Wang} which is used to generate a pseudo-random sample of $\boldsymbol{\Omega}$ for Monte Carlo integration.
\cite{Wang}'s algorithm is a pure-and-simple Gibbs sampler and easy to implement, but \cite{ONGLASSO} pointed out that it could not exactly guarantee the positive-definiteness of generated $\boldsymbol{\Omega}$. Instead they developed a positive-definiteness-assured block Gibbs sampler for Bayesian adaptive graphical LASSO.
We apply their algorithm to estimate $\boldsymbol{\Omega}$ and construct the optimal portfolio.

As \cite{Torri} mentioned, the previous studies (e.g., \cite{Demiguel2009a} and \cite{Fanetel2012}) suggested that the expected return of any asset could not be reliably estimated. Thus we will focus on the global minimum variance portfolio \eqref{gmv} which does not require any estimate of the expected return.
In performance comparison, we construct the global minimum variance portfolio of 100 assets for different sample lengths with our new approach or commonly-used non-Bayesian graphical LASSO, and compare their out-of-sample performance in long-term asset management.

The main body of this paper is organized as follows.
In Section 2, we briefly review the basic idea of graphical LASSO as well as its Bayesian interpretation, and explain the Markov chain sampling algorithm for Bayesian adaptive graphical LASSO by \cite{ONGLASSO}.
In Section 3, we report the results of experiments on long-term portfolio management with asset return data of 100 assets.
Lastly, we state our concluding remarks in Section 4.

\section{Bayesian Adaptive Graphical LASSO}
\label{model}
In this section, we introduce the basic framework of graphical LASSO and outline the Bayesian graphical LASSO approach based on \cite{ONGLASSO} which we employ for portfolio management with many assets.

Suppose $\boldsymbol{R}$ is a $n\times p$ matrix of asset return data with $p$ assets and $n$ observations and each row vector of $\boldsymbol{R}$ follows the multivariate normal distribution $\mathcal{N}(\boldsymbol{0},\boldsymbol{\Omega}^{-1})$ where $\boldsymbol{\Omega}=(\omega_{ij})$, $(1\leqq i, j\leqq p)$ is the precision matrix.
Then graphical LASSO is formulated as the following penalized maximum likelihood estimation:
\begin{equation}
\label{glasso}
\max_{\boldsymbol{\Omega}\in M^+} \frac{n}2\log|\boldsymbol{\Omega}| - \frac12\mathrm{tr}\left(\boldsymbol{S}\boldsymbol{\Omega}\right) - \lambda\Vert\boldsymbol{\Omega}\Vert_1,
\end{equation}
where $\Vert\boldsymbol{\Omega}\Vert_1 = \sum_{i\leqq j}|\omega_{ij}|$ and $M^+$ are subsets of the parameter space of $\boldsymbol{\Omega}$ in which $\boldsymbol{\Omega}$ is positive definite.
$\boldsymbol{S}$ is defined as $\boldsymbol{S} = \boldsymbol{R^{\intercal}}\boldsymbol{R}$ and called the scatter matrix.
The first two term of the objective function in \eqref{glasso} is corresponding to the log likelihood function of the multivariate normal distribution.
The last term in \eqref{glasso} is the penalty for complexity of the graph structure and $\lambda$, which is called the shrinkage parameter, dictates the magnitude of penalty for adding an extra link to the graph structure.

As \cite{Park_and_Casella} pointed out, the penalty in \eqref{glasso} is equivalent to the log density of a Laplace distribution:
\begin{equation}
\label{laplace}
p(\omega_{ij}) = \frac{\lambda}{2} e^{-\lambda|\omega_{ij}|},\quad (1\leqq i\leqq j\leqq p).
\end{equation}
Note that $\lambda/2$ can be ignored because it does not affect the solution of \eqref{glasso}.
With this interpretation, the graphical LASSO \eqref{glasso} is regarded as a maximum a posteriori (MAP) estimator.
Pushing further to this line of thinking, we can conduct a fully Bayesian analysis of the Gaussian graphical model.

In a general Bayesian framework, we first assume the probability distribution of the data generating process and set up the likelihood function. In graphical LASSO \eqref{glasso}, the data generating process is the multivariate normal distribution and the likelihood function is
\begin{equation}
\label{likelihood}
p(\boldsymbol{R}|\boldsymbol{\omega}) \propto |\boldsymbol{\omega}|^{\frac{n}{2}}\exp\left[-\frac{1}{2}\mathrm{tr}(\boldsymbol{S}\boldsymbol{\Omega})\right]\boldsymbol{1}_{M^{+}}(\boldsymbol{\Omega}),
\end{equation}
where $\boldsymbol{\omega}=\{\omega_{ij}\}_{1\leqq i\leqq j\leqq p}$ is a $p(p+1)/2\times 1$ vector of elements in the upper or lower triangular part of $\boldsymbol{\Omega}$ and $\boldsymbol{1}_{M^{+}}(\boldsymbol{\Omega})$ is an indicator function to check whether $\boldsymbol{\Omega}$ is positive definite or not.

Next, we set up the probability distribution of unknown parameters in the likelihood function, which is called the prior distribution or the prior to be short, to express non-data information about the parameters.
Although we may use the original Laplace prior for our Bayesian analysis, we instead propose to use the following prior for each element in $\boldsymbol{\Omega}$:
\begin{align}
\label{ada.prior}
 p(\omega_{ij}|\lambda_{ij}) = \begin{cases}
 \lambda_{ii} e^{-\lambda_{ii}\omega_{ii}}, & (i=j); \\
 \frac{\lambda_{ij}}2 e^{-\lambda_{ij}|\omega_{ij}|}, &  (i\ne j),
 \end{cases}
\end{align}
which means that the prior of each diagonal element is exponential while that of each off-diagonal element is Laplace.
Note that we allow the shrinkage parameter $\lambda_{ij}$ to differ from element to element.
Unlike the original graphical LASSO \eqref{glasso} where the Laplace prior is also assumed for the diagonal elements in $\boldsymbol{\Omega}$, the exponential prior is assumed in \eqref{ada.prior}.
The exponential prior is not a shrinkage prior, but this will not cause any problems because, in principle, we do no have to force the diagonal elements in $\boldsymbol{\Omega}$ to be zero because they represent links to nodes themselves and must be non-zero. Therefore, in order to achieve sparsity of $\boldsymbol{\Omega}$, we only need to apply a shrinkage prior to the off-diagonal elements\footnote{
  Alternative shrinkage priors have been developed for Bayesian graphical LASSO in the literature.
  The spike-and-slab prior (\cite{wang2015}) is a widely applied shrinkage prior in particular for variable selection in a regression model.
  Basically, it is a mixture of two distributions; one is normal with large variance and another is the Dirac delta at zero.
  When the latter is realized in the mixture of distributions, the corresponding link will be excluded from the graph structure.
  Another popular choice is the horseshoe prior (\cite{Li_Craig_Bhadra}) which assumes that each off-diagonal element in $\boldsymbol{\Omega}$ follows a half-Cauchy distribution}.

Furthermore, we treat each shrinkage parameter $\lambda_{ij}$ as an unknown parameter and assume the common prior for all $\lambda_{ij}$'s as
\begin{equation}
  \label{gamma.prior}
  p(\lambda_{ij}) = \frac{s^{r}}{\Gamma(r)}\lambda_{ij}^{r-1}e^{-s\lambda_{ij}},
\end{equation}
which is a gamma distribution.
This type of ``prior of priors'' is called the hierarchical prior.

Lastly, we derive the posterior distribution, which incorporate both data-related information in the likelihood function and non-data information in the prior, with Bayes' theorem:
\begin{equation}
\label{posterior}
\begin{split}
  p(\boldsymbol{\omega},\boldsymbol{\lambda}|\boldsymbol{R}) &\propto p(\boldsymbol{R}|\boldsymbol{\omega})p(\boldsymbol{\omega}|\boldsymbol{\lambda})p(\boldsymbol{\lambda}), \\
  p(\boldsymbol{\omega}|\boldsymbol{\lambda}) &= \prod_{1\leqq i\leqq j\leqq n}p(\omega_{ij}|\lambda_{ij}), \\
  p(\boldsymbol{\lambda}) &= \prod_{1\leqq i\leqq j\leqq n}p(\lambda_{ij}),
\end{split}
\end{equation}
where $\boldsymbol{\lambda} = \{\lambda_{ij}\}_{1\leqq i\leqq j\leqq p}$ is a $p(p+1)/2\times 1$ vector of the shrinkage parameters.
If all $\lambda_{{ij}}$'s in \eqref{ada.prior} take a common value $\lambda$ (or we may use two different values; one for the diagonal elements and another for the off-diagonal elements), the posterior distribution \eqref{posterior} is reduced to a simpler model called Bayesian graphical LASSO.
In case they can take different values for any elements in $\boldsymbol{\Omega}$ as we assume in \eqref{ada.prior}, it is called Bayesian adaptive graphical LASSO.
By adding an additional layer of uncertainty, the adaptive version of graphical LASSO can flexibly adjust the shape of the posterior distribution and may hopefully capture the reality in the financial market better.

Unfortunately, we cannot analytically evaluate the posterior distribution \eqref{posterior}, which means that we need to use a numerical approximation method to obtain an estimate of $\boldsymbol{\Omega}$.
In this paper, we employ the Markov chain sampling algorithm by \cite{ONGLASSO} for generating a pseudo-random sample of $\boldsymbol{\Omega}$, $\{\boldsymbol{\Omega^{(t)}}\}_{t=1}^{T}$, along with other parameters from the posterior distribution \eqref{posterior}, and use the generated sample in Monte Carlo integration to approximate the posterior mean of $\boldsymbol{\Omega}$ as its point estimate, i.e.,
\begin{equation}
  \label{montecarlo}
  \widehat{\boldsymbol{\Omega}} = \frac1{T}\sum_{t=1}^{T}\boldsymbol{\Omega}^{(t)}.
\end{equation}

In the rest of this section, we briefly describe the Markov chain sampling algorithm by \cite{ONGLASSO}.
In general, the term ``Markov chain sampling'' refers to an generic randam number generating method which utilizes the convergence of a Markov chain to its invariant distribution.
The basic principle of the Markov chain sampling is rather simple.
If a Markov chain is convergent to the invariant distribution, any sequence of pseudo-random numbers drawn from such a Markov chain will eventually converge to the invariant distribution.
Furthermore, under some mild conditions, the law of large numbers is applicable to the drawn sequence even though it is not an independent process.
Therefore, if we can construct a Markov chain whose invariant distribution is the posterior distribution such as \eqref{posterior}, we will obtain a pseudo-random sample of the parameters by drawing them repeatedly from the Markov chain until the sequence will be stabilized.

One of the popular algorithms for Markov chain sampling is the Gibbs sampler.
Suppose it is possible to draw each parameter in the posterior distribution from its conditional distribution given the rest of the parameters, which is call the full conditional posterior distribution in the literature.
Then the Gibbs sampler is defined as an iterative algorithm which repeatedly draws each parameter from its full conditional posterior distribution and replaces the previous value of the parameter with the new one before the next parameter will be drawn from its full conditional posterior distribution.
In this way, the new values of the parameters are obtained at the end of each cycle of the Gibbs sampler.
By construction, any sequence of pseudo-random numbers generated with the Gibbs sampler is a Markov chain whose invariant distribution is the posterior distribution, and this Markov chain will convergent to the posterior distribution in most applications including the Gaussian graphical model we study here.

To derive the Gibbs sampling algorithm for the posterior distribution \eqref{posterior}, we make use of the fact that the Laplace distribution in \eqref{ada.prior} is expressed as a scale mixture of normal distributions with the exponential distribution:
\begin{equation}
  \label{scale.mixture}
  \begin{split}
    p(\omega_{ij}|\tau_{ij}) &= \frac1{\sqrt{2\pi\tau_{ij}}}\exp\left(-\frac{\omega_{ij}^{2}}{2\tau_{ij}}\right), \\
    p(\tau_{ij}) &= \frac{\lambda_{ij}^{2}}{2}\exp\left(-\frac{\lambda_{ij}^{2}}{2}\tau_{ij}\right).
  \end{split}
\end{equation}
Define $\boldsymbol{\tau}=\{\tau_{ij}\}_{1\leqq i<j\leqq p}$.
Then the posterior distribution \eqref{posterior} is rewritten as the joint distribution of $\boldsymbol{\omega}$, $\boldsymbol{\lambda}$ and $\boldsymbol{\tau}$:
\begin{equation}
  \label{ada.posterior}
  \begin{split}
    p(\boldsymbol{\omega},\boldsymbol{\tau},\boldsymbol{\lambda}|\boldsymbol{R}) &\propto |\boldsymbol{\Omega}|^{\frac{n}2}\exp\left[-\frac12\mathrm{tr}(\boldsymbol{S}\boldsymbol{\Omega})\right]\mathbf{1}_{M^+}(\boldsymbol{\Omega}) \times \prod_{i=1}^{p}\lambda_{ii}e^{-\lambda_{ii}\omega_{ii}} \\
    &\quad \times \prod_{1\leqq i<j\leqq p}\frac1{\sqrt{2\pi\tau_{ij}}}\exp\left(-\frac{\omega_{ij}^2}{2\tau_{ij}}\right) \frac{\lambda_{ij}^2}2\exp\left(-\frac{\lambda_{ij}^2}2\tau_{ij}\right) \\
    &\quad \times \prod_{1\leqq i\leqq j\leqq p}\lambda_{ij}^{r-1}e^{-s\lambda_{ij}}.
  \end{split}
\end{equation}

It turns out that we can easily construct a Gibbs sampler for the posterior distribution in the form of \eqref{ada.posterior}.
It is straightforward to show that the full conditional posterior distribution of $1/\tau_{ij}$ $(1\leqq i < j \leqq p)$ is the inverse Gaussian distribution:
\begin{equation}
\label{fc.tau}
\frac1{\tau_{ij}} \sim \mathrm{Inverse\ Gaussian}\left(\frac{\lambda_{ij}}{|\omega_{ij}|},\lambda^2_{ij}\right),
\end{equation}
while that of $\lambda_{ij}$ $(1\leqq i \leqq j \leqq p)$ is the gamma distribution:
\begin{equation}
\label{fc.lambda}
\lambda_{ij} \sim \mathrm{Gamma}\left(r + 1, s + |\omega_{ij}|\right).
\end{equation}
To derive the full conditional posterior distribution of $\boldsymbol{\omega}$, we consider the following partition of the precision matrix $\boldsymbol{\Omega}$:
\begin{equation}
\label{partition1}
\boldsymbol{\Omega} =
\begin{bmatrix}
\boldsymbol{\Omega}_{11} & \boldsymbol{\omega}_{12} \\
\boldsymbol{\omega}_{12}^{\intercal} & \omega_{22}
\end{bmatrix},
\end{equation}
where $\boldsymbol{\Omega}_{11}$ is a $(p-1)\times (p-1)$ matrix, $\boldsymbol{\omega}_{12}$ is a $(p-1)\times 1$ vector, and $\omega_{22}$ is a scalar.
Without a loss of generality we can rearrange rows and columns of $\boldsymbol{\Omega}$ so that the lower-right corner of $\boldsymbol{\Omega}$, $\omega_{22}$, is the diagonal element to be generated from its full conditional posterior distribution.
Likewise, we can partition $\boldsymbol{S}$, $\boldsymbol{\Upsilon}$, and $\boldsymbol{\lambda}$ as
\begin{equation}
\label{partition2}
 \boldsymbol{S} =
 \begin{bmatrix}
\boldsymbol{S}_{11} & \boldsymbol{s}_{12} \\
\boldsymbol{s}_{12}^{\intercal} & s_{22}
\end{bmatrix},\quad
 \boldsymbol{\Upsilon} =
 \begin{bmatrix}
\boldsymbol{\Upsilon}_{11}  & \boldsymbol{\tau}_{12} \\
\boldsymbol{\tau}_{12}^{\intercal} & 0
\end{bmatrix},\quad
 \boldsymbol{\lambda} =
 \begin{bmatrix}
\boldsymbol{\lambda}_{12}  \\ \lambda_{22} \\
\end{bmatrix},
\end{equation}
where $\boldsymbol{\Upsilon}$ is a $p\times p$ symmetric matrix in which the off-diagonal $(i,j)$ element is $\tau_{ij}$ and all diagonal elements are equal to zero, while $\lambda_{22}$ is the element in $\boldsymbol{\lambda}$ that corresponds with the diagonal element $\omega_{22}$ in the prior distribution \eqref{ada.prior}.
According to \cite{ONGLASSO}, the full conditional posterior distribution of $\omega_{22}$ is derived as the shifted gamma distribution:
\begin{equation}
  \label{fc.omega22}
  \omega_{11} = \gamma + \boldsymbol{\omega}_{12}^{\intercal}\boldsymbol{\Omega}_{11}\boldsymbol{\omega}_{12},\quad
  \gamma \sim \mathrm{Gamma}\left(\frac{n}2+1,\ \frac{s_{22}}{2}+\lambda_{22}\right),
\end{equation}
while that of $\boldsymbol{\omega_{12}}$ is obtained as the truncated multivariate normal distribution:
\begin{equation}
  \label{fc.omega12}
  \boldsymbol{\omega}_{12} \sim \mathrm{Normal}\left(-\boldsymbol{C}\boldsymbol{s}_{12},\ \boldsymbol{C}\right)\mathbf{1}_{M_{\omega}^{+}}(\boldsymbol{\omega}_{12}),
\end{equation}
where
\begin{equation*}
  \boldsymbol{C} = \left\{(s_{22} + 2\lambda_{22})\boldsymbol{\Omega}_{11}^{-1} + \boldsymbol{D}_{\boldsymbol{\tau}}^{-1}\right\}^{-1},\quad \boldsymbol{D}_{\boldsymbol{\tau}} = \mathrm{diag}(\boldsymbol{\tau}_{12}).
\end{equation*}
and the indicator function $\mathbf{1}_{M_{\omega}^{+}}(\boldsymbol{\omega}_{12})$ implies that the domain of the distribution is truncated within
\begin{equation}
  \label{joken}
  M_{\omega}^{+} = \{\boldsymbol{\omega}_{12}:\ \omega_{22} > \boldsymbol{\omega}_{12}^{\intercal}\boldsymbol{\Omega}_{11}\boldsymbol{\omega}_{12}\}.
\end{equation}
The constraint \eqref{joken} imposed on $\boldsymbol{\omega}_{12}$ is the key to assure the positive definiteness of $\boldsymbol{\Omega}$.
\cite{ONGLASSO} suggested using the Hit-and Run algorithm to draw $\boldsymbol{\omega}_{12}$ from the truncated multivariate normal distribution \eqref{fc.omega12}.
See \cite{ONGLASSO} for more details on the derivation of their algorithm.
The outline of the Gibbs sampler is summarized as follows.
\begin{itembox}{\emph{Gibbs sampler for Bayesian adaptive graphical LASSO}}
For $i=1,\dots,p$, repeat \textit{Step 1} to \textit{Step 5}.
\begin{description}
\item[\it Step 1:] Rearrange $\boldsymbol{\Omega}$, $\boldsymbol{S}$, $\boldsymbol{\Upsilon}$, and $\boldsymbol{\lambda}$ so that $\omega_{ii}$ is in the place of $\omega_{22}$ in $\boldsymbol{\Omega}$ and partition them as in \eqref{partition1} and \eqref{partition2}.
\item[\it Step 2:] If $i \geqq 2$, $\boldsymbol{\omega}_{12}\leftarrow\mathrm{Normal}\left(-\boldsymbol{C}\boldsymbol{s}_{12},\boldsymbol{C}\right)\mathbf{1}_{M_{\omega}^{+}}(\boldsymbol{\omega}_{12})$.
\item[\it Step 3:] $\gamma\leftarrow\mathrm{Gamma}\left(\frac{n}2+1,\frac{s_{22}}{2}+\lambda_{22} \right)$, and set $\omega_{22}=\gamma+\boldsymbol{\omega}_{12}\boldsymbol{\Omega}_{11}^{-1}\boldsymbol{\omega}_{12}$.
\item[\it Step 4:] $\lambda_{12}\leftarrow\mathrm{Gamma}\left(r + 1, s + |\boldsymbol{\omega}_{12}|\right)$.
  \item[\it Step 5:] $\upsilon\leftarrow\mathrm{Inverse\ Gaussian}\left(\frac{\lambda_{12}}{|\boldsymbol{\omega}_{12}|},\lambda^2_{12}\right)$, and set $\tau_{12} = 1/\upsilon$.
\end{description}
\end{itembox}
From now on we refer to \cite{ONGLASSO}'s positive-definiteness-assured Bayesian adaptive graphical LASSO approach as Bada-PD.

\section{Performance Comparison in Long-term Portfolio Management }
\label{application}
We compare Bada-PD with non-Bayesian graphical LASSO (glasso) in terms of long-run portfolio management with the dataset of portfolios provided by Kenneth French\footnote{
  The dataset is available at Kenneth French's website \burl{http://mba.tuck.dartmouth.edu/pages/faculty/ken.french/data_library.html}.}.
Following \cite{Torri}, we choose monthly return data on 100 portfolios of US companies formed on size and book-to-market ratio.
According to the description given in Kenneth French's website, these portfolios are the intersections of 10 portfolios formed on size (market equity, ME) and 10 portfolios formed on the ratio of book equity to market equity (BE/ME).
Although \cite{Torri} used the original portfolio return data, we use the OLS residuals in the Fama-French three-factor model of these portfolio returns.
This is because the precision matrix of the original portfolio returns is not sparse, possibly due to the existence of common factors.
The data used for estimating the Fama-French three-factor model are also retrieved from Kenneth French's website.

In our empirical study, we test five scenarios: ($p, n$) = (100, 120), (100, 60), (100, 12), (100, 6) and (100, 3), which are corresponding to the sample period of 10 years, 5 years, 1 year, 2 quarters, and 1 quarter respectively.
The description of these scenarios is summarized in Table 1.
The $p / n$ ratio of the scenarios ranges from $5/6=0.833\dots$ to $100/3=33.333\dots$ and Case (b) -- (e) are corresponding to the $p > n$ problem.
In particular, Case (e) is an extreme scenario in which we construct a portfolio of 100 assets with only 3 observations.

\begin{table}[tbp]
  \caption{Descriptive Statistics of the Dataset}
  \begin{center}
  \scalebox{0.95}{
  \begin{tabular}{lcccccc}\hline
                          & p  & n & p / n & Time Period & & Data Frequency \\\hline
       Case : $ p < n $ &   &   &    &   & &\\
        (a) & 100 &  120 (10Y)  & 0.833  &  01/2001 - 09/2020 & &monthly\\ \hline
       Case : $ p > n $ &   &   &    &   & &\\
        (b) & 100 & 60 (5Y)  & 1.667  &  01/2006 - 09/2020 & &monthly\\
        (c) & 100 &  12 (1Y) & 8.333  &  01/2010 - 09/2020 & &monthly\\
        (d)  & 100 & 6 (2Q)  &  16.667 &  07/2010 - 09/2020 & &monthly\\
        (e)  & 100 & 3 (1Q)  & 33.333  &  10/2010 - 09/2020 & &monthly\\ \hline
         \end{tabular}
         }
  \end{center}
\end{table}

The backtesting for performance comparison is conducted as follows.
We form the global minimum variance portfolio \eqref{gmv} with an estimate of $\boldsymbol{\Omega}$.
For glasso, we estimate $\boldsymbol{\Omega}$ by using the functionality of \texttt{GraphicalLassoCV} in a Python package \texttt{sklearn.covariance}.
For Bada-PD, we estimate $\boldsymbol{\Omega}$ with its posterior mean\footnote{
  When optimizing a portfolio in a Bayesian approach, we generally use the predictive distribution of asset returns which is derived with the posterior distribution of the unknown parameters.
  As far as we apply the mean-variance model to portfolio selection, however, we do not have to evaluate the predictive distribution explicitly.
  As an example relevant to our study, let us consider the case of the multivariate normal distribution $\mathrm{Normal(\boldsymbol{\mu},\boldsymbol{\Sigma})}$.
  In this case, the covariance matrix of the predictive distribution is expressed as the sum of the posterior covariance matrix of the mean vector $\boldsymbol{\mu}$ and the posterior mean of the covariance matrix $\boldsymbol{\Sigma}$.
  If we assume $\boldsymbol{\mu}=\boldsymbol{0}$, the covariance matrix of the predictive distribution is identical to the posterior mean of $\boldsymbol{\Sigma}$.
  Thus evaluating the posterior mean of $\boldsymbol{\Sigma}$ or $\boldsymbol{\Omega}$ will suffice to construct the global minimum variance portfolio \eqref{gmv}.}
in \eqref{posterior}.
This posterior mean is evaluated with 10,000 draws via the Gibbs sampler by \cite{ONGLASSO} with $r=10^{-2}$ and $s=10^{-6}$ in the gamma prior \eqref{gamma.prior}.
To obtain a stable sequence of the Markov chain, we first iterate the Gibbs sampler 5,000 times as burn-in and store pseudo-random numbers drawn in the next 10,000 iterations.
In addition to glasso and Bada-PD, we form the equal weight portfolio (EW) as the benchmark. Moreover,
we try to estimate the covariance matrix with its sample analog for Case (a) in which the sample covariance matrix is non-singular since $p < n$.
The out-of-sample period is from January 2011 to December 2020 for all cases.
We examine out-sample-performance of each portfolio strategy by using a rolling window approach by rebalancing the portfolios once every three months. For the sake of simplicity, we ignore transaction fees and selling restrictions.
All computations are implemented with Python codes on a desktop PC with 128GB RAM and 3.8GHz i7-10700K Intel processor.

First, we examine 10-year performance of each portfolio strategy in terms of risk-return tradeoff.
Table 2 shows three performance measures of portfolios: mean return, standard deviation and Sharpe ratio.
As for the performance comparison among EW, glasso and Bada-PD, we focus on the Sharpe ratio in the fourth column of Table 2.
EW achieves the lowest Sharpe ratio in all cases because of its large standard deviation.
In Case (a) where $p$ is less than $n$, the Sharpe ratio is the highest for the global minimum variance portfolio with the sample covariance matrix, though it cannot be applicable to Case (b) -- (e) because the sample covariance matrix is singular in those cases.
The graphical LASSO approaches (Bada-PD and glasso) consistently out-perform EW in Case (a) -- (d) and their Sharpe ratios even exceed 1 in Case (a) -- (c).
However, the standard deviation increases sharply for both in Case (d) where the $p / n$ ratio reaches 16 and, as a results, the Sharpe ratio declines below 1.
Moreover, in Case (e) where $p / n$ ratio reaches 33, glasso fails to estimate the precision matrix.
Bada-PD, on the other hand, succeeds in estimating the precision matrix even in Case (e) and still out-performs EW in terms of the Sharpe ratio.

\begin{table}[tbp]
 \caption{Out-of-sample Performance of the Portfolios}
 \begin{center}
 \scalebox{1.1}{
\begin{threeparttable}[htbp]
\begin{tabular}{lccc}\hline
        Portfolio      &Mean Return & Standard Deviation & Sharp Ratio \\ \hline
       EW &    0.148 &  0.254 & 0.499   \\ \hline
       $p < n$  &  $   $            &                &           \\
        (a) p = 100, n = 120  &            &                &           \\
       sample covariance  & 0.299 & 0.119  &  2.332  \\
       glasso & 0.152  &0.103 & 1.265  \\
       Bada-PD & 0.155 & 0.107   & 1.246   \\ \hline
        $p > n$  &              &                &           \\
       (b) p = 100, n = 60  &              &                &           \\
       glasso  & 0.202 & 0.118 &  1.524  \\
       Bada-PD  & 0.220 &  0.137 &  1.451  \\
       (c) p = 100, n = 12  &              &                &           \\
        glasso &  0.234 &  0.145 &  1.465    \\
       Bada-PD& 0.228  & 0.145  &  1.419 \\
        (d) p = 100, n = 6  &              &                &           \\
       glasso & 0.191  & 0.210 & 0.804  \\
       Bada-PD & 0.183 &0.202   & 0.801   \\
       (e) p = 100, n = 3  &              &                &           \\
       glasso & NA \tnote{1}   & NA & NA  \\
       Bada-PD & 0.161  &  0.223  & 0.624  \\ \hline
     %\multicolumn{4}{l}{\footnotesize Note: NA means we cannot estimate by the model because of ill condition.}
    \end{tabular}
    \begin{tablenotes}
    \item[1] NA means we cannot estimate by the model because of ill condition.
    \end{tablenotes}
   \end{threeparttable}}
  \end{center}
\end{table}

Next, we compare the portfolio strategies in terms of portfolio composition such as shorting, diversification and turnover.
Following \cite{Torri}, we calculate the summary statistics for portfolio composition and report them in Table 3.
``Gross exp.'' in the second column of Table 3 is the gross exposure which is defined as the sum of the absolute values of the weights $\sum_i |w_i|$.
``Short exp.'' in the third column means the short exposure, i.e., the total amount of the short position.
``Max short'' in the fourth column indicates the maximum negative exposure of individual assets.
HDI in the fifth column is the modified Herfindahl diversification index corrected to account for short portfolio:
\begin{equation*}
\text{HDI} = \sum_{i=1}^{p} {w_{i}^{\ast}}^2,\quad w_{i}^{\ast} = \frac{w_{i}}{\sum_{i=1}^{p} |w_{i} |}.
\end{equation*}
This index measures a level of diversification of the portfolio.
``Active pos.'' in the sixth column means the percentage of active position of the portfolio.
Since we do not impose any restrictions on the weights in this study, all values are equal to 100\%.
``Turnover'' in the seventh column indicates the turnover ratio.

\begin{table}[tbp]
  \begin{center}
  \scalebox{0.85}{
  \begin{threeparttable}[htbp]
  \begin{tabular}{lcccccc}\hline
        Portfolio  & Gross exp. \tnote{2} & Short exp. \tnote{3} & Max short \tnote{4} & Active pos. & HDI\tnote{5} & Turnover \\ \hline
        EW & 1.000 & 0.000 & 0.000 & 100\%  & 0.010 & 0.000\\\hline
        $p < n$ &              &              &                &           \\
        (a) p = 100, n = 120    &              &              &                &           \\
        sample covariance & 23.195   &  11.098  & -1.626 & 100\%   &  0.017  &   11.727  \\
        glasso &5.976 & 2.488  &  -0.220  & 100\%  & 0.020 & 1.144  \\
        Bada-PD & 6.008  & 2.504  & -0.496  &  100\% & 0.021 &   1.850\\ \hline
         $p > n$ &              &              &                &           \\
        (b) p = 100, n = 60    &              &              &                &           \\
        glasso &5.471 & 2.236  & -0.224  & 100\%  & 0.021 &1.514  \\
        Bada-PD & 5.442  & 2.221  & -0.587 &  100\% & 0.021 & 1.919 \\
        (c) p = 100, n = 12    &              &              &                &           \\
        glasso &  3.935&  1.468 &  -0.768  & 100\%  & 0.023 &  3.044 \\
        Bada-PD & 3.843 &1.421  & -0.541  &  100\% & 0.022 & 2.924 \\
        (d) p = 100, n = 6    &              &              &                &           \\
       glasso & 3.223 &  1.111   & -2.019   &  100\%  & 0.077   & 4.808 \\
        Bada-PD & 2.903  & 0.952  &  -0.304 &  100\% & 0.024 & 3.330 \\
        (e) p = 100, n = 3    &              &              &                &           \\
       glasso &NA\tnote{6} & NA  & NA  & NA  & NA & NA \\
        Bada-PD & 1.801  &  0.400  & -0.099 &  100\% & 0.018 & 2.167  \\ \hline
        % \multicolumn{7}{l}{\footnotesize Notes: }\\
         %\multicolumn{7}{l}{\footnotesize (1)  Gross exp. is the sum of the absolute values of the portfolio weights $\sum_i |w_i|$.}\\
        %\multicolumn{7}{l}{\footnotesize (2)  Short exp. means is the total amount of the short position.}\\
        %\multicolumn{7}{l}{\footnotesize (3)  Max short means maximum negative exposure of each asset.}\\
        %\multicolumn{7}{l}{\footnotesize (4)$HDI = 􏰍 \sum_i {w^{\ast}}^2 $ (where $w^{\ast} = w_i /(􏰍\sum_i |w_i |)$) is modified Herfindahl diversification index. }\\
         %\multicolumn{7}{l}{\footnotesize (5) The figures are the average across all the rebalancing periods.} \\
          %\multicolumn{7}{l}{\footnotesize (6)  NA means we cannot estimate by the model because of ill condition.}\\
  \caption{Statistics of Portfolio Composition\tnote{1}}
  \end{tabular}
    \begin{tablenotes}
    \item[1] These statistics are averaged across all rebalancing periods.
    \item[2] Gross exp. is $\sum_{i=1}^{p}|w_{i}|$.
    \item[3] Short exp. is the total amount of the short position.
    \item[4] Max short is the maximum negative exposure of each asset.
    \item[5] $\text{HDI} = \sum_{i=1}^{{p}} {w_{i}^{\ast}}^2 $ where $w_{i}^{\ast} = w_{i}/(\sum_{i=1}^{p} |w_{i}|)$.
    \item[6] NA means that we fail to estimate the precision matrix because of the ill condition.
    \end{tablenotes}
   \end{threeparttable}}
  \end{center}
\end{table}

First, let us examine the risk exposures of each portfolio strategy.
For the global minimum variance portfolio with the sample covariance matrix, the gross exposure is over 23 times higher than the initial endowment and the short exposure is almost 11, in spite of the fact that the $p / n$ ratio is less than 1 in Case (a).
Thus we may conclude that using a plain sample estimate of the covariance matrix has a tendency to take extreme short positions.
For both Bada-PD and glasso, on the other hand, the gross exposure is considerably lower than that of the sample covariance case and gradually decreases as the $p / n$ ratio increases.
Note that EW takes only long position for all asset by definition and the gross exposure is always equal to 1.

Next, let us check the degree of portfolio diversification.
As for HDI, we do not observe any noticeable differences between Bada-PD and glasso except for Case (d) in which the HDI of glasso is 3 times higher than that of Bada-PD.
In Case (d), the $p / n$ ratio is beyond 16 and \emph{Max short} of glasso is much more extreme than the other cases.
We speculate that this is because the estimation procedure for the precision matrix may become unstable if the $p / n$ ratio is too high.
On the other hand, Bada-PD seems to construct portfolios with stable HDI even in Case (e) in which the $p / n$ ratio is more than 33 and glasso fails to obtain the estimate of the precision matrix.

Finally, let us look into the turnover ratio.
Obviously, the turnover ratio of EW is equal to zero and portfolios with the sample covariance matrix have the highest turnover ratio.
For both Bada-PD and glasso, the turnover ratio tends to increase as the $p / n$ ratio gets higher, though the rate of increment is less severe for Bada-PD than glasso.

\section{Conclusion}
\label{conclusion}
Limited availability of historical data on asset returns has been a hindrance to asset management because the sample covariance matrix of asset returns is singular when the number of asset $p$ exceeds the number of observations $n$.
In this paper, we explored a possible solution to this so-called $p > n$ problem in large-scale portfolio management.
To solve this problem, we proposed a new data-driven portfolio framework based on Bayesian adaptive graphical LASSO with the Markov chain sampling algorithm proposed by \cite{ONGLASSO}.
The proposed approach can directly estimate the precision matrix of asset returns even in case of $p < n$ as we demonstrated in Section 3.
We tested out-of-sample performance of the proposed approach in long-term portfolio management by using monthly return data of 100 portfolios available at Kenneth French's website in various scenarios.
In this experiment, we confirmed advantages of the proposed approach over the conventional sample covariance approach and non-Bayesian graphical LASSO in terms of return-risk tradeoff and portfolio composition.
Both Sharpe ratios and indices of portfolio composition were relatively stable for the proposed approach while they were either unstable or inestimable for non-Bayesian graphical LASSO.
Even in the most severe scenario where the precision matrix of 100 assets must be estimated with only 3 observations, the proposed approach was able to estimate the precision matrix and outperformed the equal weight portfolio without taking abnormal values in regard to indices of portfolio composition.

\section*{Acknowledgements}
This research is supported by Keio University Doctoral Student Grant-in-Aid Program.

This preprint has not undergone peer review (when applicable) or any post-submission improvements or corrections. The Version of Record of this article is published in \textit{Asia-Pacific Financial Markets}, and is available online at https://doi.org/10.1007/s10690-022-09358-8.

\section*{Ethical standards}
The author declares that the experiments in this paper comply with the current laws of Japan where we had conducted the experiment.

\section*{Conflict of interest}
The author declares that he is funded by Keio University Doctoral Student Grant-in-Aid Program in FY2020 and FY 2021 to conduct this research.
\normalsize

%glasso_app_reference→Oya and NakatsumaはJJSDに出したタイトルのもの
%glasso_app_referenceArxivold　→ arxivにある古い方のタイトル
\bibliography{glasso_app_reference}
\bibliographystyle{plainnat}

\end{document}